# PROTECT-DB: Protecting Data using Replicated State Machines: Efficient Corruption Detection & Recovery

Anant Utgikar
anant1@iitb.ac.in
CSE Department, IIT Bombay
India

S. Sudarshan
sudarsha@cse.iitb.ac.in
CSE Department, IIT Bombay
India

## Abstract

Data is critical for the operation of any organization and needs to be protected, especially against attacks that compromise the state of the database. In this paper, we explore an approach based on Byzantine-fault tolerant replicated state machines, built on top of a deterministic extension of PostgreSQL. Each replica deterministically executes transactions recorded in a shared log/blockchain. Our focus is on creating a practical system that is designed for efficient and quick detection of corruption, as well as quick repair concurrent with execution of transactions. We also present a performance study showing the efficiency and practicality of our approach. We believe our work lays the foundations for the practical use of the BFT replicated state machine approach in the context of databases.

## 1 Introduction

Data is critical for any organization, and there have been numerous incidents in recent years where attackers have broken-in and damaged data. Most databases support read/write access control but it does not protect against malicious users with administrator privileges and could compromise the contents; in fact, most applications connect to databases via logins that can update any of the application data. It could be an intruder getting administrator credentials, or due to an insider attack. In this paper, we address the problem of detection and repair of such corruption of data.

One approach is to use a set of database replicas, with the security of each node configured independently with its own administration. It is harder for the attacker to break into the security of multiple nodes, thus providing protection from corruption as long as only a minority of nodes is compromised. If the replicas are under the same administration, a compromised user account will result in the compromise of all replicas. It may be noted that replication is used to provide redundancy in the event of hardware failure, which can also be used for protection from attacks.

However, in typical implementations of database replication, updates happen at a single node and are then replicated to the other nodes. If the source of the update is compromised, data in all the replicas will also be compromised. Executing the updates independently on the database replicas is not an option since the transaction ordering may differ on different nodes, and further, some transactions may commit on one node but not on another.

Our approach is based on Byzantine-fault tolerant replicated state machines, where transactions are recorded in a shared log using consensus mechanisms, and are then executed on all replicas.

There are three major challenges in designing such a system. 1) All nodes need to agree on the order of transactions submitted to the system. 2) State of all nodes must remain synchronized while executing the same sequence of transactions. 3) Efficiently identifying and recovering replicas from corrupted / compromised state.

The first challenge, of ordering input requests, is addressed by using a shared log, through distributed consensus protocols such as Raft or Byzantine consensus protocols, or permissioned block chains (with each block containing only 1 transaction, to minimize latency). The log may alternatively be created by an external system, for example, a matching engine in a stock market, with the transactions being recorded persistently at a share depository. All database replicas fetch transactions from this shared log and execute them independently.

The second challenge, of keeping databases synchronized, arises from the concurrent execution of transactions, which can result in different serialization orders in different replicas. Serial execution avoids this problem, but is too inefficient to be used in practice. Deterministic database models [1], [23], [22] allow some amount of concurrent execution for efficiency, but guarantee that the serialization order will be the same, and hence the result as well as the final state will be the same across all replicas.

The third and final challenge, which is the focus of this paper, is how to identify compromised nodes/corrupted data, and recover from corruption efficiently, even on very large databases.

In this paper, we describe an architecture for a system designed to deal with Byzantine data corruption, with a focus on simple, efficient, and practical schemes for corruption detection and recovery.

Our specific contributions are as follows.

1. We present (in Section 2) our system model, which provides a practical architecture for building database applications, by using a BFT replicated state machine approach on top of existing databases. The model, also used in some other systems such as [28], is based on transaction code being executed at servers, while clients submit transaction requests that are basically API calls with parameters, digitally signed by the client.
2. We present our corruption detection and recovery techniques in Section 5.
   - (a) We describe two threat models (Section 5.1), one where database is corrupted but system code is safe, and another where system code also could be compromised.

(b) We discuss (in Section 5.3) two schemes to detect divergence of databases due to corruption of data. The first scheme (the passive scheme) checks the results of every user query from all replicas. The second scheme (the active scheme) explicitly invokes a special transaction to publish the database state.
(c) We propose a novel Dynamic Merkle tree (DMT) data structure which scales with size of user relation. Unlike merkle patricia tree, DMT is implemented as a relation in database, which helps leverage backup and snapshot functionality. It is created automatically for each user relation, to compare database states. To reduce Merkle root contention, we logically partition the records across multiple trees.
(d) We propose (in Section 5.4) an efficient and practical scheme for recovering from corruption, which can execute even as rest of the system continues to process latest transactions. This scheme depends on getting database snapshots, a feature that is supported by many databases such as PostgreSQL.

(3) Replicated state machines depend critically on transactions being deterministic; yet bugs can cause non-deterministic behavior, which in turn can cause database state divergence. We propose (Section 6) a simple scheme for quick detection and recovery from execution of non-deterministic transaction, which can cause serious problem if it occurs and is not detected and fixed immediately.
(4) We have implemented all our techniques, described in Sec 7, including support for clients to submit transactions and aggregating results from replicas, and support for recovery. We made extensions to PostgreSQL to support deterministic execution algorithm (presented in Section 3) based on the rw-conflict detection mechanism of Aria [23], implemented in PostgreSQL [1], and to implement Merkle trees.
(5) We present a performance study in Section 8 that demonstrates that our implementation is very efficient, and can support a throughput of over 5000 transactions per second from the YCSB benchmark even on low-end hardware. Our study shows that the overheads of deterministic execution and Merkle trees are low compared to the baseline performance of PostgreSQL.

Our approach provides a practical and effective way to detect and recover from unauthorised updates to the database. We believe it can have significant real-world impact.

Preventing unauthorised leakage of information is a separate problem and not the focus of this paper.

## 2 System Model

Figure 1 shows our system model where clients submit digitally signed transactions . These are added to a shared log, created using a consensus or other mechanisms.

Each replica in our Byzantine fault tolerance state machine replication model is implemented as an application server that forms a wrapper around a database system. The wrapper layer fetches each transaction from the shared log and performs authentication (using the signature), authorization checks. It then processes the transaction deterministically. Multiple threads enable parallel execution of transactions, with the deterministic database component responsible for ensuring ordering. The results of transaction execution at each replica are sent asynchronously to Kafka's pub-sub mechanism, to ensure fault tolerance.

We describe the components of the model in this section.

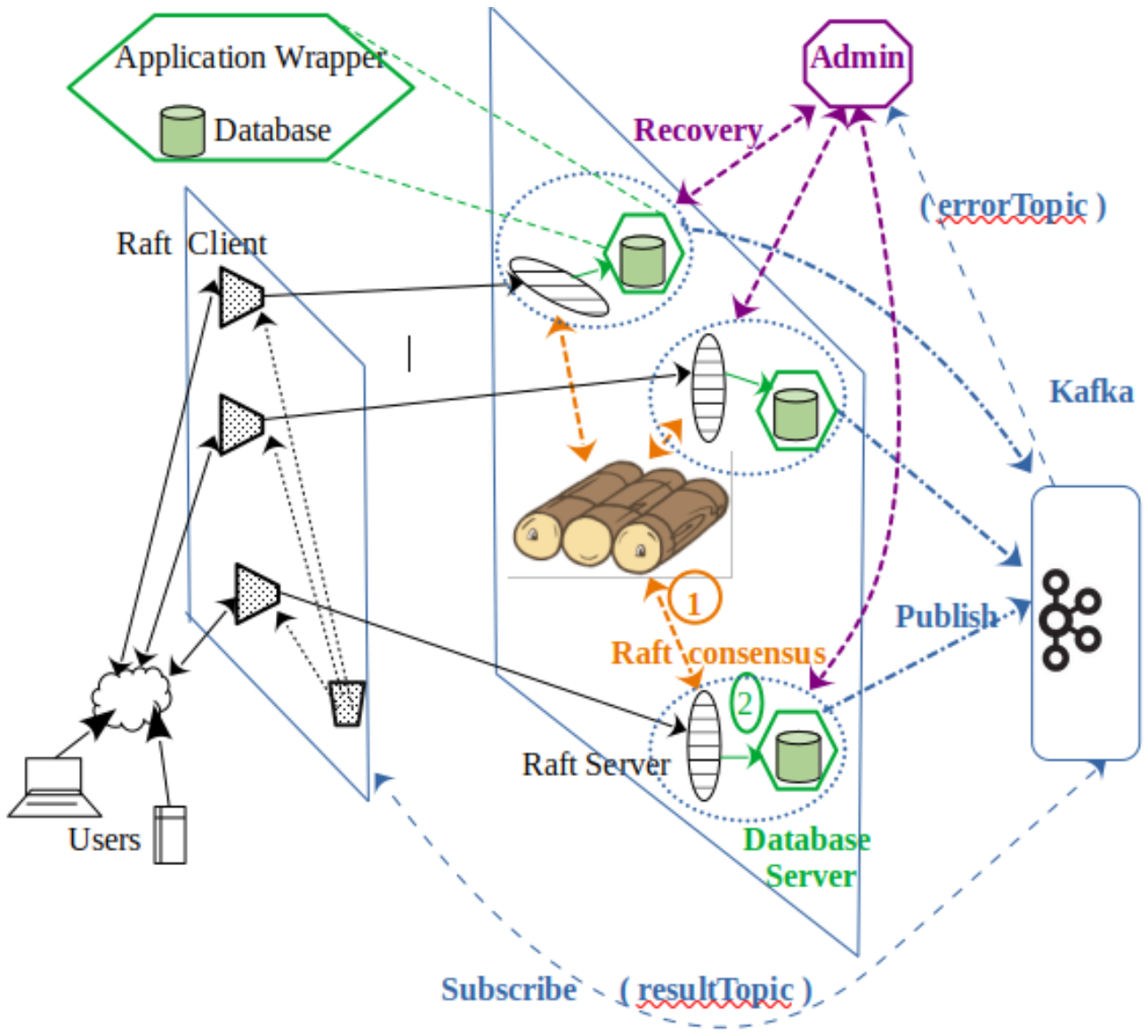

**Figure 1: System model**

### 2.1 User Interaction Model

Users submit transactions through a client, which we assume for simplicity, are trusted and not compromised. Even if they are compromised, the damage is limited to operations within the authorization scope of the specific user.

As shown in Figure 2, each transaction contains (i) the user id, (ii) a sequence number that is unique for that user (to avoid duplication/replay attacks), (iii) the users' transaction, along with its parameters, and (iv) a digital signature of the message contents.

Each Raft client also subscribes to Kafka to learn about the result of execution of the transaction at each replica. The client compares results from all nodes; as long as a majority of the full set of servers have sent the same value, it picks that value. Under the assumption that a majority of nodes are alive, accessible, and not corrupted, there would always be such a majority value, ensuring the system makes progress and does not block.

We discuss in Section 5.3 how to handle and resolve discrepancies if the results from different nodes differ. The comparison of results at the client can be optimized, by just sending a hash from all nodes except for one node that sends the full result, and comparing the hash values and validating the result against the hash.

| | | | | |
|---|---|---|---|---|
| Request | Client-ID | Sequence-Number | Transaction Statements | RSA-Signature |

**Figure 2: Request format**

[1]Available at https://github.com/zllai/AriaBC.

## 2.2 Transaction Log

Multiple users can simultaneously submit requests, with each containing one or more transactions. To ensure all replicas handle user requests in the same order, it is necessary that all replicas agree on the order of incoming requests. Thus, requests have to be added to a sequential log, which should be immutable for security reasons.

The input log can be generated by a Byzantine-fault tolerant consensus mechanism such as those used in (permissioned) blockchains, for e.g. [50]. If attacks on the consensus are not an issue of concern, and the focus is only on database state corruption, crash fault-tolerant consensus mechanisms like Raft [30] can be used, even though they are not Byzantine-fault tolerant.

There are also applications where the log is generated by an existing centralized system, but the state must be protected from attacks. For example, a stock market matching engine generates a sequential log of trades, which must then be processed by a depository which tracks who owns which shares; protecting the depository state from attack is a critical sub-problem. Where multiple parties are involved in a transaction, verification of consistency of such logs is done by periodically comparing the logs generated at the different systems, to detect and fix any inconsistencies, a process referred to as "reconciliation".

In our implementation, we create the shared log by using a open-source implementation of the RAFT consensus protocol. Clients submit transactions to a application server, which then sends the transactions to a Raft server. The Raft servers run the consensus protocol to create the shared log.

If we used blockchain-based mechanisms, the use of larger blocks can cause significant delays in transaction execution. With permissioned block chains, it is possible to use blocks containing only one or a few transactions, to avoid latency.

## 2.3 Transaction Processing at Replicas

When a transaction is committed to the log after consensus, it is sent to the application server. In the figure, it is shown as running on the same node as the Raft server, but it could be on different node. The application server performs authentication and authorization checks to determine if the user is authorized to execute the transaction. If the check passes, the transaction is sent to the underlying database which executes it. Otherwise, the transaction is deemed to have failed.

Transaction execution in our model has to be deterministic, with the serialization order identical to the order in the log. We use deterministic database execution techniques to allow concurrent execution while ensuring deterministic results, as described later. Thus, after execution, the database state of all replicas will be identical, provided the initial state was the same, and the majority of replicas are not compromised.

Once the transaction is executed, the application server sends the results to the client indirectly via a Kafka pub-sub system to handle any crashes. The results are signed by the server, and the client can verify the signature.

We will see in a later section how the corrupted nodes are handled and restored. It is possible to use the database itself as an application server, but then transactions would be restricted to stored procedures supported by the database.

## 2.4 Detecting Data Corruption

Participating server nodes that are corrupted or compromised may deliver different query results than the healthy majority, which will provide identical results.

Clients compare the results received from all the replicas. As long as a majority of the complete set of nodes have sent the same value, it is accepted, under the assumption that a majority of nodes are up and not corrupted. As long as only a minority of nodes are down or corrupted, the system will be able to make progress, else it may block. We discuss in Section 5.3 how to resolve discrepancies if the result hashes from different nodes differ.

As an optimization, only one of the nodes, chosen as the leader, can send the full result to the client, and others can simply send the result hash. The client can then compare the hashes with the locally verified hash of the full result from the leader. In case of any discrepancy, as long as a majority of the nodes sent the same hash value, the client can request one or more of the majority nodes to send the full result.

In general, a user transaction can be a combination of query and update operation. Update transactions are processed similarly to queries though their result is usually very small, indicating the status of the operation and the number of rows updated.

Our focus is on server compromise leading to data corruption at the server. However, a client may also be compromised. We note that if the client's secret key has been leaked, any transaction sent ostensibly by that client would get validated. However, the impact of such compromise would be limited to that client, with no risk to the overall system.

## 2.5 Key Management

Clients and replica server digitally sign the transaction and the execution result respectively, with their private keys. Public keys of both must be stored in the recipients' database or an external safe store. Any change in the public key should be done either with a certificate or through a trusted symmetric channel. User keys being stored in the database, any update to keys is also a transaction on the blockchain. Thus, if a request signed with the old key arrives after the key-update request in the blockchain, it will be rejected.

# 3 Deterministic Execution Algorithm

Several deterministic execution schemes e.g. [1] allow transactions to execute concurrently, but commit serially. Aria [23], Harmony [22] are deterministic execution algorithms, where multiple transactions are batched and executed concurrently, on the same database snapshot. If a conflict between read-write sets is detected within a batch, the lower ID transaction is allowed to commit. Conflicting higher ID transactions are run again in the next batch, on a newer snapshot. The order is guaranteed to be the same on all replicas, although it may not match the given input order.

Our deterministic execution algorithm pseudocode is shown in Fig 1, and is based on implementation of AriaBC [23] on PostgreSQL [2]. Each worker thread picks the next transaction ($T_k$) from the input log and executes the Run_Tx function. The function atomically performs two operations: (a) takes the current database snapshot $S$ and (b) records the lowest unfinished transaction-id as *compare_from_id*

[2]Available at https://github.com/zllai/AriaBC.

(the last committed transaction ID + 1) (*Lines* 25 – 26). (We assume that transactions are numbered sequentially in log/commit order.) Executing atomically ensures that all updates from transaction-IDs prior to *compare_from_id* are committed in the snapshot $S$. Thus, updates of $T_k$ need to be checked for any possible conflict with transactions that had started but had not finished, before $T_k$ started.

As noted in [23], the read-write check also detects phantom conflicts since PostgreSQL uses leaf conflicts for checking conflicts between predicates and updates/deletes/inserts, and these conflicts show up as read-write conflicts on index leaves.

**Algorithm 1** Deterministic Transaction Execution Algorithm

```
 1: RW_SET[] setA, setB                          ▷ global read-write hash table sets
 2: RW_SET mapToTable(Tx_id);                    ▷ based on switchThreshold
 3: last_committed_tx_id ← 0                     ▷ global across threads
 4: compare_from_id ← 0                          ▷ private to worker
 5:
 6: function Commit (Tx_id_i)
 7:     currentSet ← mapToTable(Tx_id_i)
 8:     (currentSet != mapToTable(Tx_id_{i-1})) ? empty(currentSet) : NO-OP
 9:     Insert (i, rw_set_i ) into currentSet
10:     last_committed_tx_id ← Tx_id_i
11: end function
12:
13: function CheckConflict (TX_ID1, TX_ID2, RW_SET_i)
14:     for ( i ← TX_ID1 ; (i ≠ TX_ID2) ; i ← i + 1 ) do
15:         currentSet ← mapToTable ( i )
16:         if ( (RW_SET_i conflicts with RW_SET_j from currentSet ) &
                (TX_ID_j >= compare_from_id ) ) then return true
17:         end if
18:     end for
19: return false                                 ▷ No conflicts
20: end function
21:
22: function Run_Tx (T_k) :
23:     for ( loop ← true ; (loop) ; ) do
24:         oldest_commit_id ← last_committed_tx_id
25:         compare_from_id ← oldest_commit_id + 1
26:         rw_set_k ← Execute(T_k, DB_Snapshot() )
27:         while (oldest_commit_id ≠ (k − 1)) do
28:             WaitForCommitSignals()           ▷ periodic or count-based etc
29:             latest_commit_id ← last_committed_tx_id
30:             loop ← CheckConflict( oldest_commit_id,
                              latest_commit_id, rw_set_k)
31:             oldest_commit_id ← latest_commit_id
32:             ( loop ) ? break : continue
33:         end while
34:     end for
35:     Commit(T_k)                              ▷ Publish read-write sets
36:     CommitSignalBroadcast()                  ▷ Signal waiting workers
37: end function
```

A worker executes $T_k$ on the database snapshot $S$, to determine the read-write-set which was not known prior to scheduling $T_k$ (*Line* 26). They are committed after If all preceding transactions have committed before snapshot $S$ was taken (the check in the while loop in *Line* 27), $T_k$ can commit updates. Otherwise, if at least one preceding transaction had not committed before snapshot $S$ was taken, $T_k$ will not commit, as in [22], [23]. It waits to check for data conflicts with preceding transactions through their notification (commit signal) of read-write sets published. This wait can be either for a certain time period, or a count of commit signals, etc (*Line* 28).

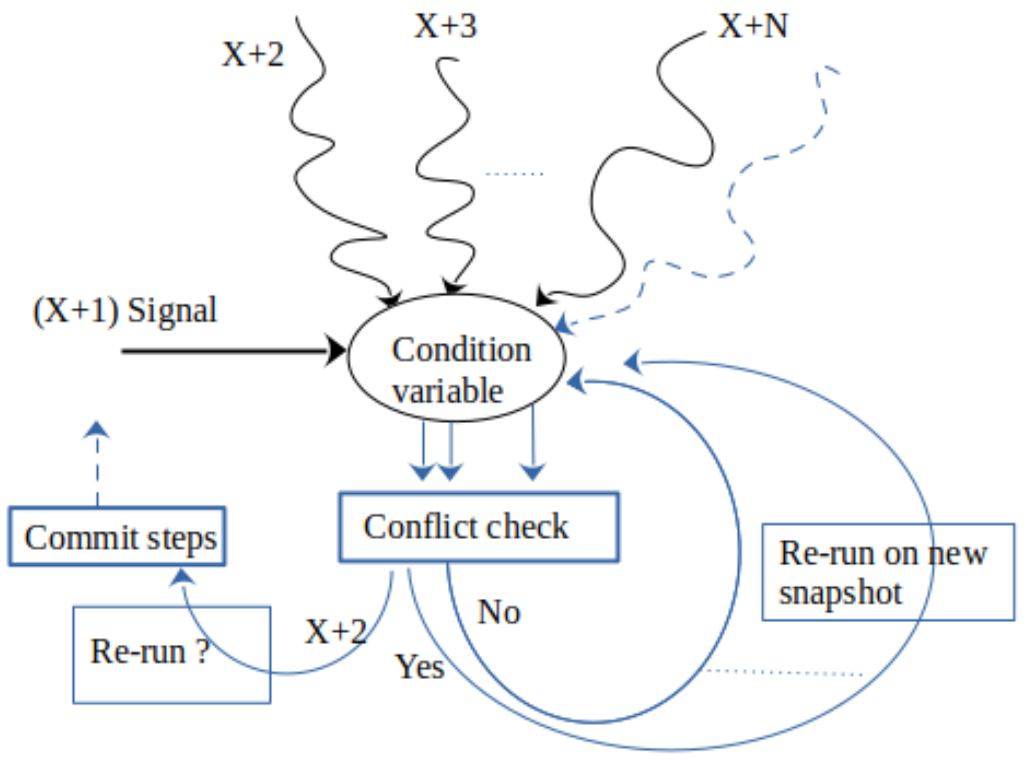

**Figure 3: Workers and condition variable**

A commit signal indicates the availability of a read-write set from a preceding transaction that committed. When a commit signal is received, the worker will (a) record the ID of that transaction and (b) compare the read-write set of $T_k$ with all those who have published a commit signal (i.e. committed, *Line* 30). In the optimized case, this translates to the number of signals reaching threshold count or timer expiry. It then compares the read-write sets of all transactions till this recorded ID, with its own read-write set (*Line* 13).

If there is a conflict, $T_k$ is restarted by taking a new database snapshot with the latest state after a waiting time as determined in *Line* 28. The intuition behind the condition variable wait and conflict checking code in the while loop above is depicted pictorially in Figure 3. Once conflict checks with all (if any) preceding transactions are passed, $T_k$ can commit (*Line* 6) its changes and save read-write sets to the global hash table (*Line* 7 – 10) and then send a broadcast signal to other workers (*Line* 37).

A hash table stores read-write sets of moving window of committed transactions. Once a transaction $T_z$ is old enough that no currently executing transaction started before $T_z$ committed, its read-write sets can be deleted from the hash table.

To simplify deletion of data from the hash table, we use a pair of hash tables $P$ and $Q$, where one table is written for a threshold count of transactions, and then the other table is used, again till the threshold. The threshold is chosen larger than the maximum number of worker threads $N$, so that when $Q$ becomes full, all current transactions in $P$ are redundant can be deleted, and symmetrically for $P$ becoming full. Due to commit ordering, with $N$ threads, the oldest executing transaction cannot be more than $N$ transactions before the newest currently executing one.

Upon re-execution on a new snapshot, the transaction may follow a different codepath from the original simulated execution, resulting in new read-write sets. This is expected to be a rare scenario, and even in this case, the new execution will account for all conflicts.

## 4 Dynamic Merkle Tree

As the database size increases with row count, more tuples map to a leaf node in the Merkle tree. Hence, during corrupt replica recovery, the time to fetch the tuples and compare them across the replicas increases linearly with the number of tuples mapping to a merkle leaf node. To reduce the recovery time of a corrupt replica, we propose to limit the number of tuples mapping to a leaf node, with a Dynamic Merkle tree data structure that scales dynamically as the user relation grows/shrinks.

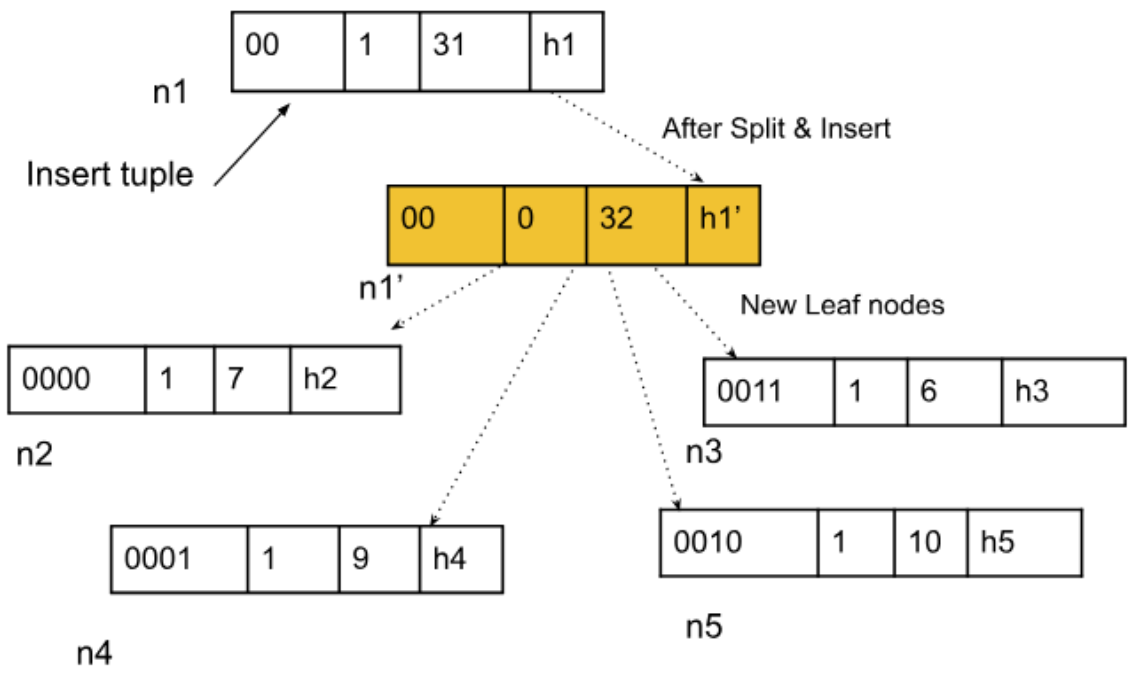

**Figure 4: 4-way Leaf Node Split in Dynamic Merkle Tree**

While the structure of B+tree also limits number of keys per node and scales dynamically with input values, it is non deterministic. Thus, two replicas executing transactions in the same logical order, but with some non-conflicting operations ordered differently, can internally have two different B+tree structures for the same data. Thus it is not possible to extend a B+tree with node hashing to efficiently find differences between replicas.

We require the ability to compare replicas where tuple insertions may occur in different orders and thus, internal tree structure must be determined only by the tree content, regardless of the order of operations, Our proposed Dynamic Merkle tree data structure preserves this property of determinism, while allowing the tree to grow or shrink incrementally ensuring that the number of tuples mapping to a leaf node is bounded. When the number of tuples mapping to a leaf node exceeds a certain threshold, our Dynamic Merkle tree is restructured by making that node as internal node and creating multiple children among which the mapping tuples are divided based on tuple hash.

Depending on the skew in input data, some tree branch may grow deeper by adding child nodes more often than some other branch, resulting in a tree with non-uniform depth from root to leaf, across branches. Thus, unlike a B-tree, a Dynamic Merkle tree is not a balanced data structure. However, as long as hashing functions are well designed, leaf depths will be relatively uniform.

Similarly, when tuples are deleted and the count of all tuples mapping to descendant nodes of an internal node falls below a certain threshold, the child leaf nodes can be merged into the parent node. Note that node splitting and merge-up can happen recursively across multiple levels until the aggregated tuple count does not violate the respective threshold.

Another important requirement is the ability to read snapshots of the Merkle tree , which are used for comparing the state across nodes. Since many databases already support snapshots of database data, we can reuse existing database snapshot support by storing the merkle tree as a relation, as explained earlier.

In Section 4.1, we describe the algorithms used to update merkle tree during insert, delete and update operations on user relation. In Section 4.2, we describe the data structures maintained within database to support resizing the merkle tree dynamically.

## 4.1 Dynamic merkle tree tuple operations

For all types of operations that modify a user relation tuple, the first step is a common procedure to identify the merkle tree leaf node to which tuple maps. For this, we traverse the merkle tree down from the root node, at each step, finding the child node whose *node_id* is a prefix of the tuple_hash. At the root node, the prefix length used for comparison is say $k$ and the length is incremented by the same amount $k$ at each level down the tree. As discussed earlier, whenever a tuple in user relation is modified, the tuple hash determines the merkle leaf node the tuple maps to.

*4.1.1 Dynamic merkle tree tuple insertion .* When a new tuple is inserted in the user relation, the leaf node to which the tuple maps in the merkle tree, is determined as described above. For all nodes along the path from the root to the leaf, the entries in the *R_merkle_nodes* relation are updated by (a) incrementing the tuple_count and (b) including the tuple_hash in the node_hash.

When inserting a tuple in the user relation, if the merkle leaf node it maps to, has *tuple_count* equal to a certain threshold, it is split by creating new leaf nodes as its children, thus making the original leaf node, an internal node. The tuples mapping to the original leaf node are distributed between its child leaf nodes based on which new *node_id* their tuple hash prefix matches. The B+tree index on tuple_hash, helps efficiently search for the tuple hashes mapping to a node in the merkle tree, since range search accesses contiguous memory, rather than random I/O.

As shown in Figure 4, the leaf node $n1$ has 31 tuples mapping to it in a merkle tree configured with $k = 2$ and threshold of 31. For simplicity, common parameter partition_id is omitted, and is identical for all parent and child nodes. After the new tuple insertion, the maximum threshold leaf size would be exceeded, leading to node split. The leaf node $n1$ (before split) becomes parent node $n1'$ (shown shaded in figure) with 4 child leaf nodes (in this 4-way case) shown as nodes $n2$ , $n3$ , $n4$ , $n5$ respectively. The tuples are mapped to child nodes based on their hash prefix and may not be equally distributed among the new child leaf nodes. In the example, after tuple insertion, node $n1'$ with *node_id* (00) matches prefix of all 32 tuple hashes, however, the number of tuple hash prefixes matching new leaf nodes are 7, 9, 10 and 6 respectively.

For newly created leaf nodes, it is necessary to compute the hash values from the tuple hashes, since the earlier leaf node $n1$ only had aggregated hash for all tuples. Note that the *node_hash* value of the node $n1$ is only updated to incorporate hash of tuple being inserted, since the union of set of tuples mapping to its new child nodes is the same as the set of tuples that mapped to it before the split.

*4.1.2 Dynamic merkle tree tuple deletion .* Similar to insertion operation, when a tuple is deleted from the user relation, the leaf node to which the tuple maps in the merkle tree, is found as described earlier. For all nodes along the path from the root to the mapped leaf, the entries in the *R_merkle_nodes* relation are updated by (a) decrementing tuple_count and (b) excluding the tuple_hash from the node_hash.

Conversely to discussion of tuple insertion, we first check the parent of leaf node to which the deleted tuple mapped. If the number of tuples mapping to the parent node of that leaf node falls below a certain threshold, all the child leaf nodes are merged into parent node thereby reducing the height of that branch of the tree. The parent node (in *R_merkle_nodes* relation) is rewritten as a leaf_node by merging tuples at all the child nodes. The hash of the parent node needs to be recomputed only for the tuple being deleted, since all the other tuple hashes already mapped to it through the earlier child leaf nodes. This merging-up may have to be repeated for parent of the parent node up the tree, if *tuple_count* of all peers of the parent node (other children of parent of the parent node) is 0. This merging-up is performed recursively until we reach a parent node whose tuple_count is above the threshold. The example in Figure 4, can be viewed as an example for tuple deletion from user relation, by merkle tree node-merging with direction of arrows reversed.

*4.1.3 Dynamic merkle tree tuple update .* Similar to insertion and deletion operations, when a tuple in the user relation is updated, the leaf nodes to which old and new tuples map in the merkle tree, is found as described earlier. A tuple update operation results in merkle tree update equivalent to old tuple being deleted and new tuple being inserted.

Note that when both old and new tuple_hashes map to same merkle partition, the *tuple_count* will not change for any node which is common parent of leaf nodes to which the old and new tuple map. However, it may cause split along the branch leading to leaf of new tuple hash and shrink/merge along the branch leading to leaf of old tuple hash . In the special case where both old and new tuple_hashes map to same leaf node, the *tuple_count* will not change for any node. Also, no change is needed in the merkle tree structure since no merkle tree nodes are created or deleted. In this case, it is possible to optimize the *node_hash* operations by combining into single operation if the hash-combining function is associative and distributive.

## 4.2 Dynamic Merkle Tree Data Structure

For each user relation R, a new relation *R_merkle_nodes* is created internally with an entry for each merkle tree node (both leaf and internal) having attributes as follows:

*( part_id, node_id, is_leaf, tuple_count, node_hash )*

Here, the first two columns indicate the merkle partition_id (*part_id*) and the *node_id* which is unique within the partition. An internal node has $2^k$ child nodes whose *node_ids* are $k$-bit longer than their parent, and are generated by suffixing parent *node_id* with each of ($2^k$) bit sequences . For example, with $k = 2$, a parent node has 4 children, the *node_ids* for child nodes are generated by suffixing one of {00, 01, 10, 11} to the *node_id* of the parent node.

Note that the *node_ids* of child nodes of an internal node are mutually exclusive and completely span the universe defined by the *node_id* of the parent node. Thus, if a tuple mapped to the parent node, it will map to exactly one child node as only one child *node_id* will match the the bit-prefix of tuple_hash. Matching recursively , we locate the leaf node to which the tuple maps. A tuple mapping to a leaf node is considered to be mapping to all parent nodes of the leaf node, uptil root node. A *node_id* matches bit-prefix of the hash of all tuples mapping to that node. Thus, when a tuple is updated, its tuple_hash will change, and the updated tuple may map to another leaf node, different from the original leaf node.

The value of the next attribute *is_leaf* indicates whether the node is internal (0) or leaf (1). The last two attributes respectively indicate (a) the count of all tuples mapping to this node and (b) the aggregation of tuple hashes of all tuples mapping to this node. A tuple hash is computed as the hash value of all the attributes of the row. A B+tree index is created on tuple_hash, which is created as a covering index with extra attribute of user relation key resulting in hash values that are contiguous.

The merkle tree schema presented above remains the same for both static and dynamic merkle trees. For static merkle tree, the number of partitions, nodes and node-types (*is_leaf*) remains constant irrespective of the number of tuples . In contrast for dynamic merkle tree, only the number of partitions remains constant. Leaf nodes with too many mapping tuples are split to create child leaf nodes, thus changing node-type of original node to internal. In both cases, *tuple_count* and *node_hash* change with tuple operations.

# 5 Corruption Detection & Recovery

In this section, we discuss different threat models and the modes of detection of a corrupt replica. We present a mechanism to update the Merkle tree with each transaction, along with partitioning. Finally, we describe how to recover a compromised node and discuss integrity constraint violations during recovery.

## 5.1 Threat Models

We consider two threat models reflecting minor and major levels of security breaches.

In the first threat model, we assume the database system code is not compromised. Even when an attacker has broken into the system and tampered with data, the database configuration and automatic update of the Merkle tree remain intact. On the compromised node, the Merkle tree will generate a different root hash that does not match that of a healthy node. When root hashes of all nodes are compared, compromised nodes can be identified as long as they form a minority.

In the second (or major) threat model, the database system code may also be compromised; thus, the Merkle tree hash may not match the underlying data, and the node may fool a recovering node into accepting wrong results. Yet, the result of the transaction provided by this replica based on tampered data will not match the majority result, thus indicating a security breach. To handle this model, we need to first restore the code, then recompute Merkle trees, and then restart the protocol.

Activities such as user public key update are done by the system administrator. A compromise of the system administrator's private key could impact multiple users. We assume in this paper that the administrator credentials are not compromised. The rest of this paper focuses on the first threat model.

## 5.2 Fault Models

The basic model considers that the input log is not compromised and only protection of database is necessary. In the case that input log needs to be protected, byzantine consensus protocol (like PBFT, hotstuff) can be used instead of Raft. To protect against the result aggregator being compromised, signed results from all replicas can be forwarded to the user client. The client can verify each signature and find the majority result.

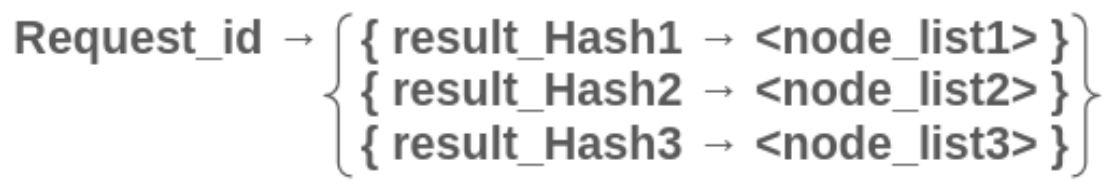

**Figure 5: Transaction Result Aggregation Map**

## 5.3 Corrupt Replica Detection

All nodes receive an identical sequence of globally ordered transactions. In passive detection, the result of each transaction is monitored. If the result from any two database replicas differs, at least one of them must be corrupted. Since only a minority of nodes may be compromised, we can determine the correct transaction result based on the majority. However, data corruption may manifest in transaction result mismatch after a very long time. Hence, a system administrator may want to detect data corruption early, rather than passively waiting for it. We propose a passive and an active scheme below, and later present a mechanism to recover a corrupt replica.

*5.3.1 Passive Detection Using Transaction Result .* When the result of a transaction execution is found to be different on different replicas, we can infer that one or more replicas have been corrupted.

Each Raft client maintains a data structure shown in Figure 5, which stores for each transaction (request) identifies the set of result hashes received from the replicas, along with the list of replicas that returned that value. This data structure is used to determine the correct value (returned by a majority of the nodes in the system). All nodes whose result hashes differ from the majority are detected as having been corrupted.

This passive detection scheme only detects corruption that affects the result of a query. If some data that has been corrupted is not queried for a long time, that corruption would correspondingly not be detected until it is queried.

*5.3.2 Active Detection using Merkle Tree .* In the active detection scheme, the database state at all replicas at corresponding points in the input log are compared with each other, by using a Merkle tree whose root node hash represents the state of the entire database (or root hashes of all trees in the partitioned case). The database state is obtained by taking a database snapshot, and all queries, including queries on Merkle tree data, run on this snapshot. A textbook description of Merkle trees for such comparison may be found in, e.g., [40]

To actively detect a corrupt node, a special transaction, 'compareStates', is inserted into the blockchain to gather database state at all replicas at the same instant in the transaction log. When a compareStates transaction is found, the application server at the replica does the following atomically: (a) it saves a database snapshot to be used later during the recovery procedure and (b) it publishes the root hash of the Merkle tree, as the state. These states are then compared, and any divergence is used to detect corrupt replicas.

Such active detection can be performed either periodically by the system or initiated on demand by system administrator.

Some database replicas may be unavailable or lagging in execution, and the root hash at that instant may not be available within a reasonable time. Hence, comparison of the rest of the nodes can be done without it, possibly after a timeout. As long as healthy nodes form the majority, we can check the health of such lagging nodes when the root hash at that point in the log becomes available.

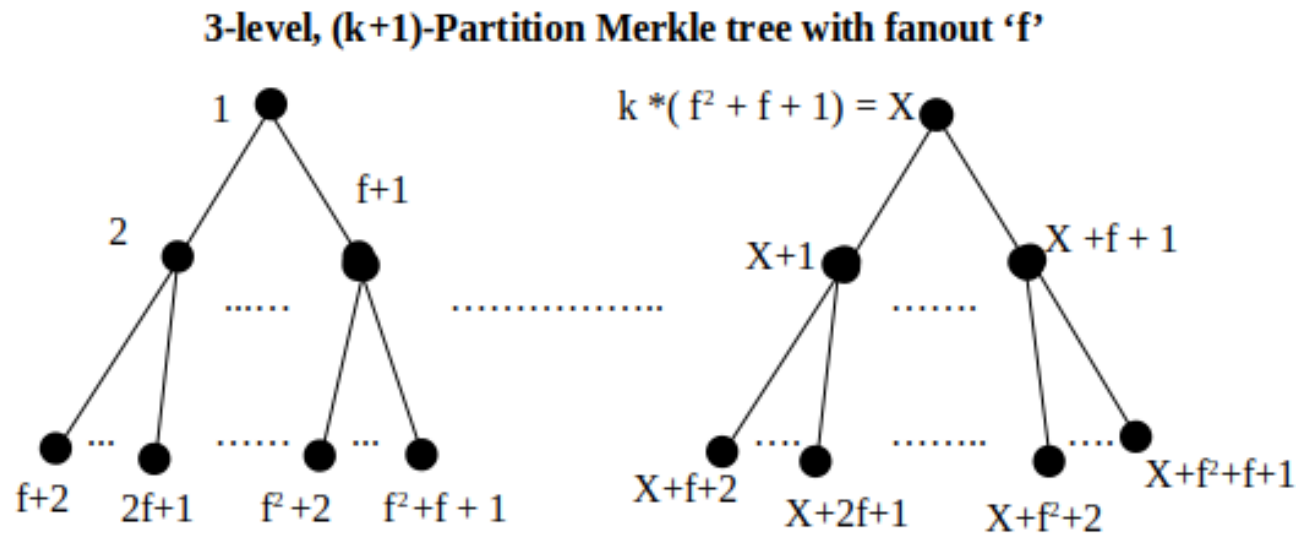

**Figure 6: 3-level, (k+1) Partitioned Merkle tree**

*5.3.3 Merkle Tree Update.* Each time a transaction updates a row in a user relation, the value of Merkle tree leaf to which it maps (mapping function), is updated. This leaf update involves removing the contribution of the previous data hash and incorporating the new data hash. This triggers a value update at parent of leaf node, which is done similarly. This updating of parent node value continues recursively up the tree till the root node. All these updates of Merkle tree nodes are done as part of the original transaction itself, which updated the row.

We note that the Merkle tree nodes are maintained as records of a system relation, ensuring existing conflict detection mechanisms also handle conflicts on the Merkle trees.

When merkle trees are not maintained, two or more concurrent transactions updating distinct rows in a user relation can execute independently by locking their respective write-sets. With merkle tree maintained, if the modified rows map to the same leaf node, write-sets will imply conflict at all nodes from leaf to root. Even if the modified rows map to different leaf nodes, both transactions will update all common parent nodes, which will always include the Merkle tree root. Thus, write-sets of such transactions will not be disjoint if both write to a common Merkle tree.

**Partitioned Merkle Tree:** Conflicts in updating common items in write-sets, in particular a common root node, is a severe bottleneck for transaction concurrency. To reduce conflicts, we logically partition each relation and compute a (smaller) Merkle tree on each partition [14]. A transaction updating a Merkle tree leaf now causes updates to its parent nodes only till the root node of its partition. Thus, write-set conflicts between transactions when updating the Merkle tree are significantly reduced.

Figure 6 shows a partitioned tree, with the node ID assignment that keeps node IDs unique across all partitions. Node hash data of all nodes across all partitions of the tree is maintained in a single relation, whose rows are identified by the node ID.

Transactions modifying user data that map to leaf nodes in the same partition may still conflict while updating Merkle tree nodes, but the partitioning scheme and count can be chosen such that conflicts are greatly reduced. Transactions updating data in different partitions will have no common Merkle tree node, and such updates can be applied independently. On the other hand, a transaction updating data across multiple partitions will limit the concurrency gain offered by partitioning. In all cases, the transactions commit in the same global input log order defined by the blockchain.

**Algorithm 2** Recovery Alorithm

1: **procedure** RECOVERY (REPLICA $N_k$)
2: ▷ **active detection start**
3: mark_log_offset ← current_log_offset ('compareStates' transaction)
4: For each replica $N_i$:
5: $S_i \leftarrow$ *DB_Snapshot*(replica $N_i$ , mark_log_offset)
6: $M_i \leftarrow Merkle_Tree(S_i)$
7: $[G_{majority}, G_{rest}] \leftarrow Group_replicas_by_merkle_root_hash()$
8: ▷ **active detection completed**
9: ▷ **recovery phase start**
10: $(N_{ref}, M_{ref}) \leftarrow$ reference instance of $(replica_ids, G_{majority})$
11: *Compare_merkle_trees_and_Replicate_rows* $(M_k, M_{ref}, N_k, N_{ref})$
12: *Replay_input_tx_logs_and_rejoin_cluster*(mark_log_offset)
13: ▷ **recovery phase complete**
14: **end procedure**

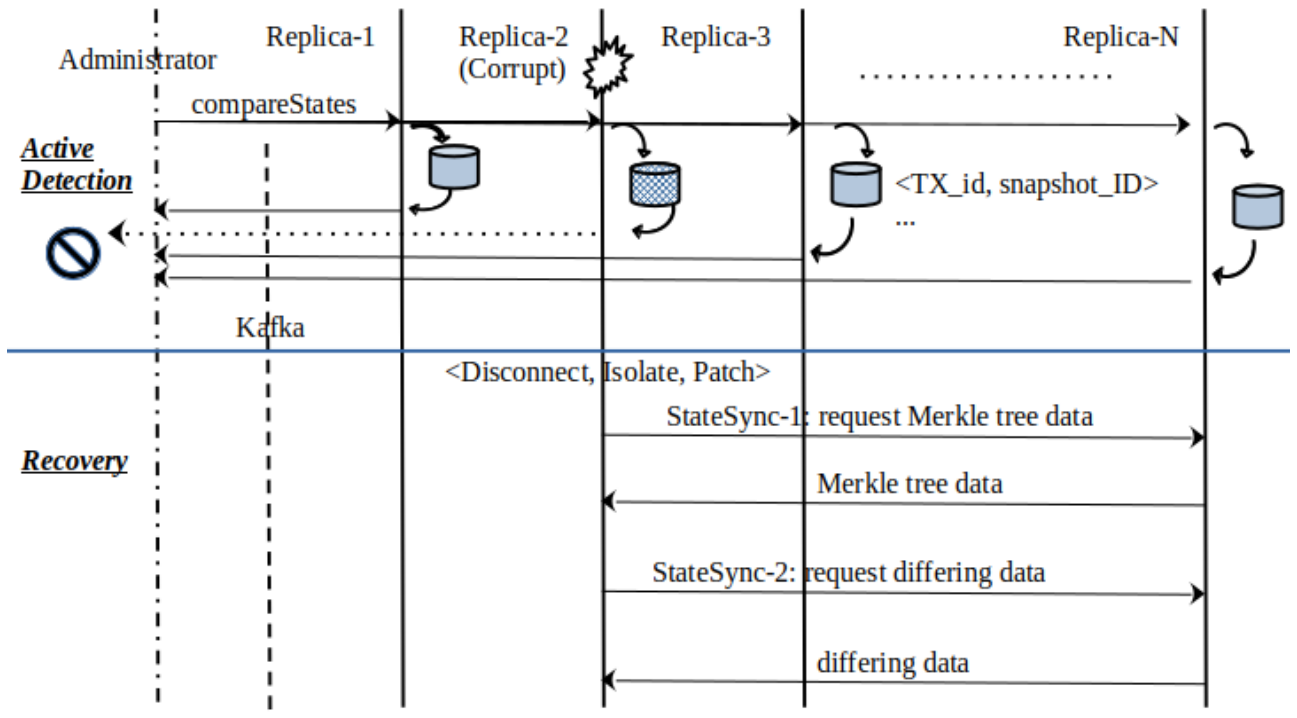

**Figure 7: Procedure to recover a failed replica**

When the content of a database replica is compromised, as identified by a difference from other replicas in either Merkle tree root hash or a transaction result, that indicates the need for a recovery mechanism to bring the replica back to operational state.

## 5.4 Corrupt Replica Recovery

The state of a database replica may be corrupted by malicious updates. System failures, like power or network, or component failure, which result in a replica falling substantially behind the rest of the system can also be treated similar to corruption, to recover the state, without playing back a very long transaction log.

Before initiating recovery, we first stop processing at the corrupted replica and ensure that the intruder no longer has access to the system. With manual intervention, the security loophole on the corrupted replica is fixed, like a security patch or system update, to prevent it from being compromised again in the future.

We describe the procedure performed by the system administrator to recover a corrupted replica pictorially in Figure 7, and with pseudocode in Algorithm 2.

The first step is to identify a healthy replica to use as a reference. To do so, a logical snapshot $S_1$ of the current database state is taken at all nodes (Line 4,5), as of a certain point in the transaction log as explained in Section 5.3.2. The logical snapshot $S_1$ includes the user data as well as corresponding Merkle trees. The default mechanism for taking snapshots in PostgreSQL takes care of this step.

We choose one of the majority of replicas which have an identical Merkle tree root, thus an identical database state, as the reference replica (Line 10).

Next, we identify the differences between the Merkle trees of the corrupted replica and the reference replica (Line 11) on the snapshot $S_1$ taken in the earlier step. Since the root hash of the corrupted replica is different, the hash of at least one of the children of the root node must differ from that in the reference replica. By recursively comparing the hashes of children down the tree, we can identify leaf-nodes whose hashes do not match across replicas. In case of partitions, as proposed in the previous section, the comparison must be done between all corresponding partition-trees.

From mismatching leaf-nodes, we identify the differing rows in the database. The differing rows are then resolved by replicating the state of those rows from snapshot $S_1$ of a reference database replica.

At the end of the above phase the database has been recovered up to the point in input transaction log when the snapshot was taken. The state of the erstwhile corrupted replica is then brought to the current state by replaying the input transaction log starting from just after the point in the log where the snapshot was taken, during the active detection procedure (Line 12).

It is possible to restore a minority of corrupted replicas concurrently with this procedure, even if they were corrupted at different times. Also, a healthy replica may serve as a reference for multiple corrupt replicas simultaneously.

Our mechanism utilizes snapshots, which most current generation databases support. In PostgreSQL, we can get a specific snapshot with an identifier, and then run multiple queries on that snapshot. We use this feature in our implementation.

*5.4.1 Integrity Constraint Discussion.* The database state at the healthy replicas was reached by executing transactions in the order they appeared on the blockchain. However, the recovery process imports all the differences from the healthy reference replica to the corrupt replica, without any order. Hence, while the recovery is being run, the corrupt database instance may temporarily violate integrity constraints. For example, a relation P may refer to a foreign key in relation Q. A transaction may add new entries to both relations on a healthy replica. However, the recovery may first update relation P entirely, while the foreign key entry in relation Q is yet to be updated.

Hence, constraint checks are disabled while the recovery is in progress and re-enabled when complete. After all differences are applied, the recovered replicas will satisfy all constraints, and integrity checks are re-enabled.

# 6 Handling Non-Deterministic Transactions

The architecture of our system assumes that transactions are deterministic. However, a badly written transaction may be non-deterministic; for example, the execution result may depend on a random number, or dynamic system-parameters like time, thread-id, etc. In this case, despite executing transactions in the same deterministic order across all replicas, database states across replicas could diverge. This is different from database corruption due to a security breach, which was discussed earlier, and could result in each replica returning a different result, or having a different state after the transaction.

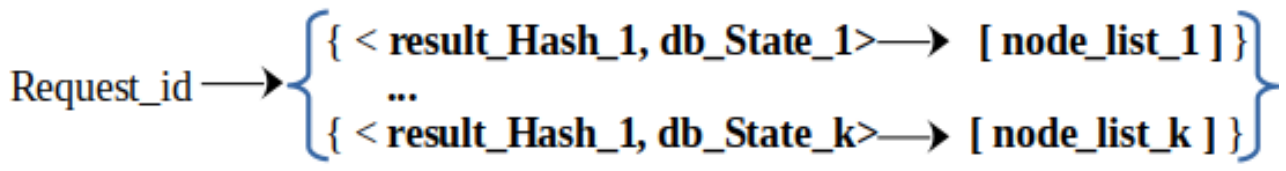

**Figure 8: Transaction Result and State Aggregation Map**

Our system already compares the query results from different replicas, and will detect the case where the query result is non-deterministic. However, transactions may update the database non-deterministically, resulting in divergent states across replicas, while returning a deterministic query result.

We present below detection and recovery mechanisms for such non-deterministic transactions.

## 6.1 Detecting Nondeterminism

To detect database state mismatch across replicas due to a non-deterministic transaction, we propose a simple extension of the scheme used earlier for passive detection of data corruption.

Writes to the database only return an execution status, and even if the actual writes are different in different replicas, the status may be identical. To detect divergence, in addition to returning transaction results, each replica now also sends the updated Merkle tree root hash of all the partitions that are updated, for each updated relation. The partition root hashes are compared, along with result hashes, to detect divergence.

Figure 8 shows an extension of the Figure 5 for detecting database state divergence. The client collects the result hash and updated partition hash information (i.e. partition ID and root hash) for all partitions updated by the transaction. Note that this extended scheme can also be used to improve accuracy of passive corruption detection described earlier, in Section 5.3.1.

The above detection mechanism is likely to be particularly useful for debugging transactions, since it catches non-determinism immediately after the transaction commit, allowing easy detection of the cause for non-determinism. In a well designed system there should ideally be no occurrence of non-determinism, but the check could be done anyway since it is very low overhead, similar to assert() function calls in programs.

However, in case non-deterministic execution does happen in a production system, it is important to be able to recover quickly from it. We describe how to do so next.

## 6.2 Recovery process

When a database state divergence is detected across replicas, it could be either because of a security breach or non-determinism. While the underlying cause for the non-deterministic execution of the transaction needs to be fixed, if it was a rare error, example due to a “heisenbug”, there may not be another occurrence for a while, giving time to fix the bug. A repeated error would, however, cause slowdown of the system due to recovery, and need urgent fixing.

As long as a majority of the nodes give the same result, the recovery method described in Section 5.4 can be used without any change. However, it is possible that there is no majority group. In that case, recovery can be initiated after deciding which replica to use as the correct copy, which can be done by either choosing one which is in a majority, or if there is no majority, by a simple majority, or by decision of a system administrator.

The recovery mechanism proposed earlier for corruption can be applied here with minor enhancements as follows:(a) how reference replica is chosen based on Merkle root hash (or in case of partitioned tree, partition root hash) values across replicas and (b) recovering multiple replicas (in the limiting case, for all replicas but one).

In case some replicas report common states, we could use the set of replicas with highest cardinality (if one exists) as reference. As observed earlier, in the extreme case, each replica may give a different root hash. In this case, there is no single source of truth and we can choose any one as reference. Those replicas whose database states are different from reference replicas, are referred hereafter as diverged replicas.

A special transaction 'saveSyncState' is inserted into the shared input log, listing all reference replicas. It causes each replica to save database snapshot and check if it is listed as reference replica. If a replica is not a reference, it needs to stop processing and recover state from one of the reference replica listed, and after that it can fetch newer requests from the shared input log. The replica listed as reference can simply save the snapshot, and continue to process newer requests without interruption.

Note that non-determinism is detected after transactions have committed at the replicas. Transactions added to the shared input log before the special recovery transaction is inserted, may have a dependency on the non-deterministic transaction. Since our detection mechanism is quick and no manual intervention is necessary, there will be very few such transactions. Database state changes due to these transactions, as well as those due to the non-deterministic transaction, are replicated from a reference replica and would be consistent after recovery. But results returned to users based on these transactions could potentially have errors.

If a transaction does not depend on the non-deterministic update, the majority would have returned the same result as the reference value. If it did, and there is a majority with the same value for the non-deterministic result, the dependent transaction would have returned the same result on all these nodes. If there is no majority, the dependent results would also not be in a majority except in an unlikely scenario where divergent inputs gave the same transaction result for the dependent transaction. In this unlikely case alone we may need to inform a user that the value returned may changed.

# 7 Implementation

We use Raft for crash-fault tolerant consensus, and Kafka for fault-tolerant asynchronous messaging. In this section, we describe several implementation details.

## 7.1 Transaction flow through System

Transaction requests are sent by user to Application server which then sends them to a Raft server. Raft leader establishes consensus with other server instances to determine a shared order, in a crash fault tolerant manner. In our implementation, each Raft server replica has a unique ID and is configured and initialized identically with a Raft server instance, an application server, and a PostgreSQL database. The Raft server instance on each node sends transactions that have reached commit status on the Raft log to the application server. The server is multi-threaded, allowing concurrent execution of multiple transactions (deterministic ordering is handled at the database side). Database requests are sent to the local PostgreSQL database server through the connection pool. The number of threads and size of the pool is configured to be the same as the number of workers on the PostgreSQL server.

The application server receives query results from the PostgreSQL server, and on completing the transaction, publishes the transaction result to Apache Kafka result topic [13] along with its server ID . During corruption recovery, Administrator connects to application server to (a) initiate Merkle tree comparison (b) lookup data items using functional index and (c) replay transaction log.

Kafka service has separate topics to hold transaction results, and server errors, respectively. The Raft client which received a user transaction gathers results for the transaction from all replicas via the Kafka result topic and stores them in a hash-map as described earlier. If the client detects a divergence in results from one or more replicas it publishes a message to a Kafka error topic for the Administrator to trigger recovery action. NuRaft [29] is an open source implementation of Raft.

## 7.2 Database processing

Database executes transactions deterministically as per our algorithm, which is based on the implementation of Aria [23] in Postgres, provided by the authors of HarmonyBC [22]. [3]

*7.2.1 Initialization .* We create schema-specific user relations, functions on each Postgres database replica, with all constraints as deferrable . All user relations and Merkle trees are set to the same initial state across replicas.

Rows in a user relation are mapped uniformly with a hash function to Merkle leaf nodes . The hash of each user data row is computed by hashing attribute values with Blake3. Hash can be computed from tuple attributes, using any standard hashing function like Blake3 [31], MD5 [36],[45], SHA-512 etc. Tuple Hash used to map user relation tuples to merkle leaf nodes, could either be the entire hash or a fixed subset of it (like a prefix or suffix or intermediate range), as computed by hashing function on all or some subset of attributes, as long as same mechanism is used for entire merkle tree, across all replicas. Note that the hash length is chosen to be long enough so as to differentiate among tuples and it is not required to offer strong cryptographic security. Tuple hash could be implemented in Postgres either (a) as a new generated column in the user relation or (b) as a function index.

Hashes of all rows mapping to a leaf node are combined using the XOR function to allow efficient incremental update of the leaf hash values. Hash-value of a parent node is computed by combining the hash values of all child nodes again using the XOR function.

*7.2.2 Merkle Tree Relation .* For each user relation, we create a Merkle tree, implemented as an index. The Merkle tree is configured identically on all replicas (schema, partitions, thresholds, fanout, mapping and hashing functions, etc). We created a new index type in PostgreSQL to maintain the Merkle tree, and is implemented as an internal relation. For each update done to the user relation by a transaction, the index update is invoked automatically by the database code, which in our solution updates the corresponding rows of the Merkle tree relation. An optimization is to keep the internal merkle tree nodes in memory, so that the transaction updates are fast, and do not need to be written to disk each time. The merkle tree could be computed at startup. Normal concurrency control and recovery techniques get applied on the Merkle tree relation.

For partitioned Merkle trees, one relation holds the hash values of Merkle nodes of all partitions, each with a different ID range. With $P$ partitions, fanout of $F$ and $k$ levels in the Merkle tree, there will be $P$ root nodes, and node count $N = (F^{k+1} - 1)/(F - 1)$ per partition and $P * F^k$ total leaf nodes and Merkle relation with $P * N$ rows. We vary the number of partitions from 100 to 1000 in order to bring down the data conflicts between transactions by orders of magnitude.

*7.2.3 Dynamic Restructuring .* As discussed earlier, a fixed size merkle tree is not the most efficient to recover databases of different sizes. We scale the merkle tree dynamically with the user relation, keeping tuples per leaf node bounded within a range. The lower bound of the range is the merge threshold whereas the split threshold is the upper bound. It should not be too small (causes frequent merkle tree restructuring) nor too large (becomes more like static merkle tree). When splitting a large leaf node to create child leaf nodes, we can utilize additional bits from the tuple_hash, to distribute the mapping tuples to new child leaf nodes.

## 7.3 Corruption recovery

The system administrator inserts a special 'compareStates' transaction into the input log, to compare the database state at all replicas at the same instant in the transaction log, say $P_{tx}$ , which saves snapshot $S_1$ at each replica. $S_1$ includes all database relations, including the user relation and the corresponding Merkle trees.

The recovery program simultaneously connects to reference replica $N_{ref}$ with a read-only session and to the corrupt replica with a read-write session. On the corrupt replica, commands are run to (a) disable new connections, (b) disconnect all connections other than the recovery session, and (c) disable / defer the constraints. On both replicas, (a) the database snapshot is set to $S_1$ and (b) Merkle node hashes are requested (of all partitions in partitioned case). Merkle trees at the corrupted replica and the reference replica $N_{ref}$ are compared to find all leaf nodes whose hashes differ.

From mismatching leaf-nodes, we identify differing rows in the database by reverse-lookup using functional index mapping Merkle tree leaf nodes to the user relation rows on snapshot $S_1$. Such an index is easily implemented in PostgreSQL using a "function index" on the table, wherein an index is created with its key as a function. . The state of the erstwhile corrupted replica is then brought to the current state by replaying the input transaction log starting from the point $P_{tx}$ where the snapshot $S_1$ was taken. This is very efficient except in case the corruption is major, when it is best to copy over a snapshot without Merkle tree comparison.

A small fanout merkle tree is more levels deeper, compared to a merkle tree with larger fanout, which is shorter. When recovering a corrupt replica from a healthy replica, respective nodes of merkle tree are compared starting with root node and then all next level children of nodes with differing *node_hash*. For a smaller fanout merkle tree, this will require many roundtrip messages, accessing merkle node information from memory and then sending it over the network. With other parameters constant, this latency is significantly reduced for merkle tree with larger fanout.

[3]Available at https://github.com/zllai/AriaBC.

*7.3.1 Corruption Recovery optimizations.* As described earlier, in our proposed merkle tree based approach to corruption recovery, majority of time is spent in (a) fetching the merkle trees from disk and (b) comparing across replicas across network and (c) fetching the rows which map to differing merkle leaf nodes, from disk and finally (d) updating the corrupt data on disk.

We optimize the fetching of merkle trees from disk in (a), by creating an index on the merkle relation attributes used for search. A merkle tree node is uniquely identified in the merkle relation with triplet attributes (nodeid, prefixlen,) . To batch the lookup of multiple merkle nodes at one level, IN query is not efficient for long list of triplet parameters. Hence, a temporary table is created from arrays of triplet parameters using unnest. To reduce the latency further in (b), we use asynchronous futures and connection pool to concurrently fetch the merkle node data from both replicas. For quick lookup in (c) we use a covering index with user relation key on top of merkle tree hash index. Finally for (d), we use batching to restore multiple rows of data to fix the corrupt replica.

The comparison of merkle trees described earlier, compares merkle nodes across replicas at one-level, finds differing nodes and then repeats for their next-level child nodes. This can be optimized by comparing all nodes which are higher-order descendents (multiple level down) of differing nodes. This is especially helpful when the merkle tree fanout is small (e.g. 4 or 8) or network latency between replicas is high, and comparable to disk read latency. However, for co-located replicas, with large merkle tree fanout (say 32), fetching 2-level data (1024 nodes) take much longer (800ms) compared to two one-level lookups of differing nodes (150ms each). For taller merkle trees with small fanout, the lookup overheads make multi-level lookups more efficient.

## 8 Performance Evaluation

We compare the performance of ProtectDB (our proposed model) on local and cloud setups, on the popular OLTP benchmark, YCSB. Our local setup has nodes with AMD Ryzen7 5700G having 16 cores, NVMe SSD, and 16GB RAM running 64-bit Ubuntu 22.04 LTS. The AWS setup has *m6id.4xlarge* EC2 instances, with 16 vcpus, Ubuntu 24.04 and NVMe SSD storage with Linux kernel version 6.14.0-1018-aws and GCC version 13.3.0

### 8.1 Evaluation on single node

We first conduct an ablation study of deterministic transaction processing on a single machine.

*8.1.1 Impact of determinism.* We compare ProtectDB with the baseline of (non-deterministic) Postgres with Serializable transaction isolation, on the local setup. As seen in Figure 9, the throughput initially increases linearly with number of threads. For higher number of threads (greater than 11), memory IO becomes a limiting factor, and beyond, the contention between threads leads to a slight dip in throughput. ProtectDB gives a peak performance which is nearly 25-30% lower than that of native PostgreSQL in each case - with and without merkle tree maintained.

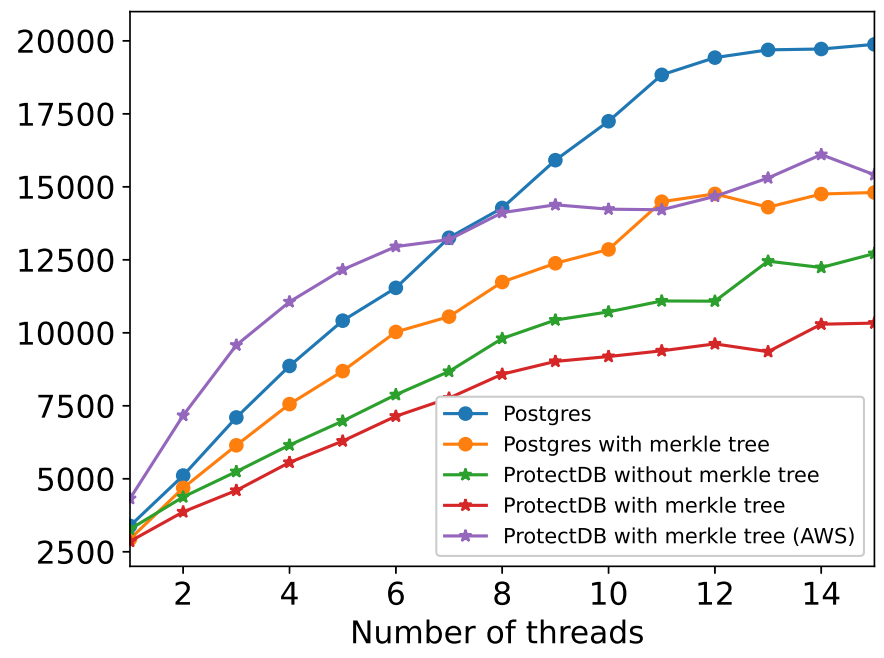

Performance with passive detection: proposed deterministic model and postgres: with and without merkle trees

**Figure 9: Impact of Determinism and merkle trees on YCSB-A**

*8.1.2 Impact of maintaining Merkle tree.* For each user relation (e.g. usertable in YCSB), a Merkle tree is constructed with 100 partitions and fanout of 16, with 3M tuples. As seen in Figure 9, throughput increases linearly and later saturates in both cases, with and without Merkle tree. ProtectDB with merkle tree, gives a peak performance of around 11000 transactions per second, which is nearly 15-20% lower than the case without merkle trees, for the same number of concurrent threads considered.

*8.1.3 Impact of merkle tree configuration .* We evaluated performance of deterministic execution with varying number of merkle tree partitions (50-500) and fanout (8-128) and saw no noticeable difference in performance. Higher configurations are intuitively expected to help reduce conflicts between transactions. However, for system throughput already saturated at concurrency (memory IO limit) less than number of partitions, no gain results. Impact of transaction conflicts was evaluated with workloads that update (a) same merkle leaf and (b) leafs with common parent node and (c) leafs within one merkle partition. Experiments showed throughput impact of 50% and 60% in the two cases respectively.

We also evaluated memory usage by the merkle index for YCSB user relation with 3M tuples which occupied 155MB in postgres, with its primary index being 64MB. Using 8-byte tuple-hash and 83 bytes per merkle tree node, we evaluated merkle index size for range of fanout, split threshold and partition count. This excludes the lookup index defined on user relation and used to search for keys mapping to a merkle leaf node, and requires around 90MB. Merkle tree is configured with partition count = 100, split threshold = 32 and fanout = 32, unless the parameter is specified as the variable in each ablation study. As seen in Table 1, when the split threshold is small, the internal nodes split very frequently and create far more empty nodes, when fanout is larger than split threshold. This causes merkle index size to be larger than the user relation itself. This effect reduces since the split threshold reduces as fanout increases.

As seen in Table 2, when fanout is small, a lean and tall merkle tree is formed, compared to the case with large fanout. As fanout increases, merkle tree depth decreases and the index size generally increases as each split adds large number of underfull (tuple count < 25% of split threshold). This holds unless the fanout increases beyond split threshold, ensuring fewer underfull nodes are created.

| Split Threshold | 4 | 8 | 16 | 32 | 64 |
|---|---|---|---|---|---|
| Merkle size (MB) | 407 | 377 | 374 | 111 | 12 |

**Table 1: Merkle Index size for YCSB 3M tuples**

| Fanout | 4 | 8 | 16 | 32 | 64 |
|---|---|---|---|---|---|
| Merkle size (MB) | 27 | 52 | 49 | 111 | 47 |
| Depth | 6 | 4 | 3 | 3 | 2 |
| Node count | 247k | 468k | 437k | 990k | 416k |
| Underfull (<25%) Nodes | 33k | 226k | 226k | 885k | 226k |

**Table 2: Merkle Index size for YCSB 3M tuples**

| Partition count | 100 | 200 | 300 | 400 | 500 |
|---|---|---|---|---|---|
| Merkle size (MB) | 111 | 23 | 36 | 47 | 59 |
| Node count | 990k | 212k | 317k | 423k | 528k |
| Underfull (<25%) Nodes | 885k | 5k | 75k | 225k | 390k |

**Table 3: Merkle Index size for YCSB 3M tuples**

As seen in Table 3, when partition count increases, merkle tree size generally increases, similar to trend and reason seen with increasing fanout. This remains true unless increase in partition count reduces node occupancy to just below split threshold, thereby reducing the number of underfull nodes drastically.

## 8.2 Corruption Recovery performance

Corruption recovery is evaluated by hosting healthy and corrupt databases across replicas. In rest of this section, we first compare our recovery approach with other solutions and later evaluate impact of different parameters and optimizations on recovery.

*8.2.1 Other Corruption Recovery Approaches.* We evaluate our proposed corruption recovery mechanism using YCSB benchmark with 1, 3 and 5 million tuples (244 MB, 750 MB and 1221 MB respectively). We compare the performance with snapshotting technique described in ChainifyDb [38]. Note that the compared ChainifyDb [38] approach does not have provision to selectively restore the modified tuples, unlike our merkle tree based approach. Hence, the measurements for the ChainifyDb [38] approach, is corresponding to complete backup and restore. The recovery time measurements of our merkle tree approach correspond to the case where 30 merkle leaf nodes are corrupted. Our approach detects corruption as soon as results from majority replicas are published, which is almost instantaneous. Thus recovery happens much sooner, and completes in a fraction of the time of [38] approach and many order of magnitude better than disk dump approach. Recovery with the naive approach to save snapshot to disk and restore-from disk file performs 3x worse than ChainifyDb [38] for all sizes.

*8.2.2 Corruption Recovery Ablation Study .* We evaluate the impact of different factors like (a) recovery optimizations described in Section 7.3.1 and (b) network delay between replicas. We use network delays as observed on AWS, namely, same availability zone network delay is under 1ms, same region delay is 1-2ms and neighboring regions is 10-50ms, as seen in [10].

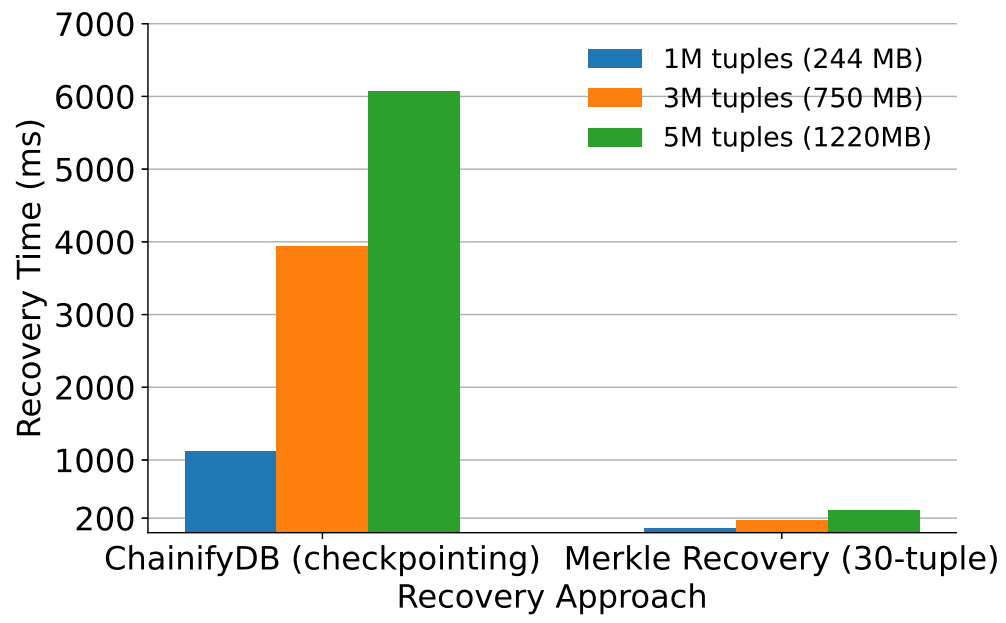

**Figure 10: Comparison of Chainifydb [38] recovery with ProtectDB for YCSB**

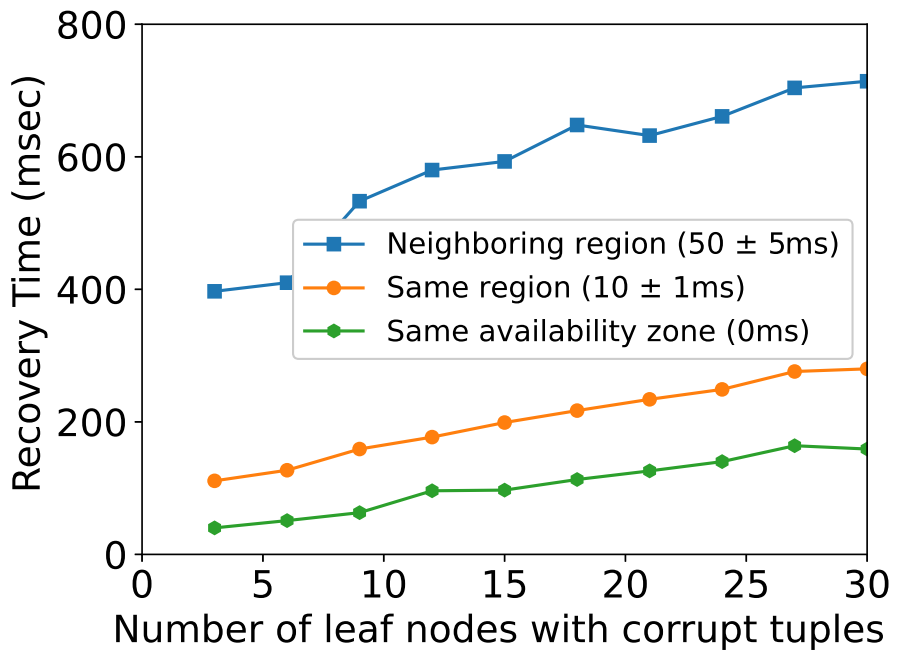

**Figure 11: Database Corruption Recovery across replicas with network delay (fanout 32) for YCSB**

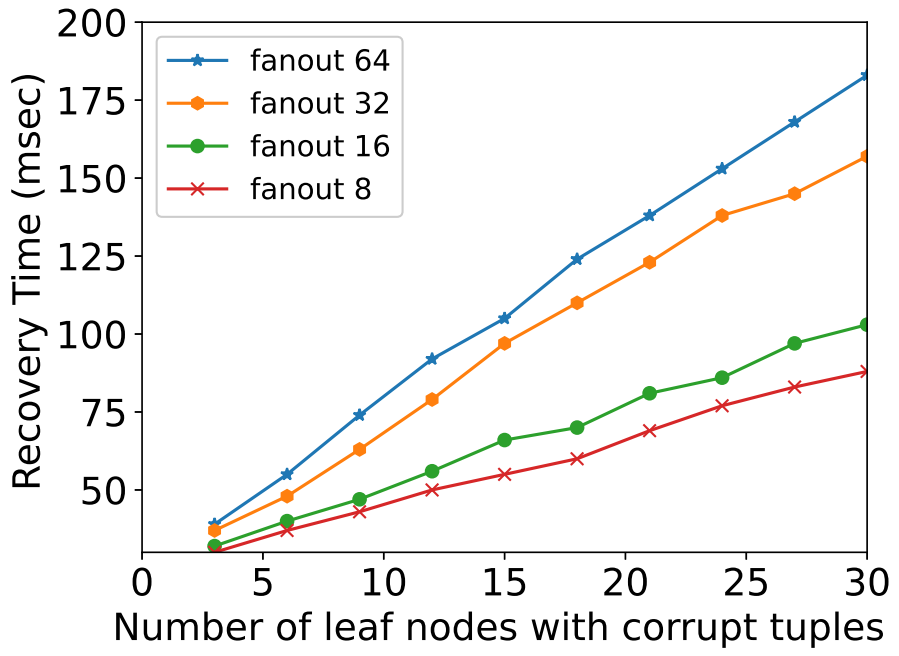

**Figure 12: Database Corruption Recovery across replicas over local network (different fanout) for YCSB**

Optimization (a) improves recovery time from quadratic to linear in number of corrupt tuples. The recovery time improves nearly 35% with optimization (b), than without. Optimization (c) significantly improves recovery time in number of corrupt tuples. Optimization (d) marginally improves recovery time in number of corrupt tuples.

To simulate corruption of $k$ leaf nodes, $k$ corrupted tuples are generated randomly. For small $k$, the corrupted tuples get distributed randomly among the leaf nodes, over all partitions. This measures the worst recovery performance since, it requires us to explore path to each leaf node from its partition-root. Note that even if one corrupted tuple maps to a Merkle leaf node, all user relation rows mapping to that leaf node need to be read from disk and compared, to determine if it is corrupted and needs to be restored.

**Figure 13: Database Corruption Recovery for different partitions**

For example, Merkle tree with 2-levels and fanout of 32 has 1024 leaf nodes per partition. With 100 partitions and user relation of 3M tuples, nearly 30 tuples map to a leaf node. The user relation tuples are corrupted through insert, delete and update operations which are equally likely.

Figure 12 shows the recovery time of database using our approach with varying fanout. Observe that the recovery time increases as fanout increases since a mismatching parent node requires more child nodes to be fetched and compared. As seen in Figure 11, for small corruption, the recovery is sub-second, and time increases linearly with number of leaf nodes corrupted. However, since our approach only copies corrupted tuples, the recovery time is much less than the time for copying the entire data (which could take hours or days on very large data) as is done by some earlier systems [38].

### 8.3 Discussion

When a healthy replica processes transactions and concurrently serves as a reference for recovering a corrupt replica, its throughput has negligible impact (nearly 2% on 10650 tps). However the corrupt replica sees 35-40% slower recovery (1470msec vs 1060msec for 300 tuple corruption) with busy healthy replica than with idle replica.

Note that our corruption detection mechanism is very quick (within few transaction execution time) and thus, recovery has to deal with very few state differences and is accomplished within few milliseconds. Also, once a corrupted replica has been synchronized to the state of the snapshot of a healthy replica, it fetches newer transactions from shared consensus log at its own pace, without impacting any other replica in the system.

The overhead of maintaining a snapshot and executing transactions on it, is measured. We evaluate the performance impact on the system while the recovery of a faulty node is in progress.

## 9 Related Work

We discuss related work on Byzantine-fault tolerant replicated state machines in Section 9.1, on Recovery in Section 9.3, on deterministic execution, in Section 9.4. and other related work in Section 9.5.

### 9.1 Byzantine Fault Tolerance

Byzantine-fault tolerant replicated state machines keep data safe from attacks as long as a small minority (usually less than 1/3rd) are corrupted [8], [12].

In a replicated state machine, a common input log of transactions is read by each of multiple replicas, and processed deterministically in the input log order. The log could be generated by a crash tolerant consensus protocol like Raft [30], or by a Byzantine-fault tolerant consensus protocol, such as those used in (permissioned) blockchains, or by any other application dependent mechanism. Most BFT replication solutions focused on non-database applications and process the log serially in each replica. For general database applications, serial execution has a very negative impact on throughput. Our work focuses on protecting the database state, given a sequential input log of transactions.

Most of the work on BFT for data storage systems does not support SQL, and none of them addresses efficient corruption detection and recovery, which is very important for large databases. Basil [41] supports Byzantine fault tolerance, but on a key-value store that supports limited operations, Callinicos [32] also supports key-value stores with limited operations ("mini transactions") and does not support SQL, while Janus [53] addresses asynchronous BFT. Augustus [33] presents a scalable storage system, supporting partitioning and replication, which is robust in the presence of Byzantine failures, but has restrictions on transactions. None of the above systems support SQL, nor do they discuss corruption detection and recovery, which is the focus of our paper.

Early work in BFT replication for databases [15], [17], assumed serial execution. HRDB [47] allows concurrent execution on SQL databases but forces replicas to follow the same order as defined by a primary database. HRDB depends on strict 2PL, whereas current generation databases often use (serializable) snapshot isolation. Byzantium [34] executes transactions in three parts: begin, execution and commit. Begin and commit operations execute serially in the log order. Transactions can execute concurrently, on a snapshot determined at start. Mitra [24] allows a coordinator to run transactions concurrently first, and then replicas follow the same order. Both these approaches support SQL queries. Pesto [42] extends the ideas behind Basil to support SQL, but requires significant changes to the database. However, none of the these approaches addresses corruption detection or recovery.

HyperLedger Fabric [4] uses an execute-order paradigm, whereas our transaction model is similar to order-then-execute of [28], but neither [4] nor [28] considers corruption detection and recovery.

### 9.2 Merkle tree based approaches

ChainifyDB [38] addresses SQL database recovery based on database snapshots, which are very impractical for large databases both for creation and for copying over a network. (Their experiments were on 1,000,000 records which is relatively small.) In contrast we use Merkle trees for efficient recovery. Furthermore, they use triggers for tracking changes; an intruder who can modify the database can equally well disable triggers when modifying the database. In contrast, we use index update mechanisms to maintain Merkle trees, avoiding this vulnerability.

Layered storage techniques like LMPT [9] and SWMT [39] are agnostic of consensus, execution and do not address detection or recovery from corruption. Credereum [20] traces state change to user transactions but does not provide solution for recovery.

BitTorrent version 2 (BEP52) [7] also uses merkle tree structure to identify corrupted blocks of a download and relies on block retransmission to recover. Libra [3] and Jellyfish merkle tree in Diem [54], used optimized binary merkle tree for storage but neither of them presented a scheme to recover a corrupted participant node. Many merkle tree variants [37] address proof of membership and some are more expensive to maintain like Verkle trees [21] whereas our focus is on efficient comparison of sets for recovery.

## 9.3 Database Recovery

ARIES [26] recovers crashed relational databases using checkpoints and applying undo, redo operations. [25] restores database state from previous checkpoint and then replays the transactions.

Various replication techniques are discussed in [16], [46] like (a) active model or state-machine replication that we use and (b) passive or primary-backup and (c) chain replication.

Multiple alternatives for recovery [19] are (a) lazy replication to reduce overhead on functional replicas and break down the recovery of a replica outdated from long time, into incremental phases (b) full vs incremental data copy, and (c) reconstruction table proposal which tracks ID of last transaction that updates an entry. In case of partitioned database [18], leader replica of each partition may be different, which executes transactions with strict 2PL and recovery may be performed independently for each partition.

## 9.4 Deterministic Databases

Deterministic databases e.g. Calvin [43], and [35] execute transactions in a way that the serialization order respects the given input order, while executing concurrently to improve throughput [2]. They address deterministic execution on a single node as well as deterministic execution in a partitioned-parallel database system. Sentry [48] detects latent data corruption but does not provide solution to recover.

Conflict detection in Aria [23] and Harmony [22] re-executes conflicting transactions in the next batch. The order is guaranteed to be the same on all replicas, although it may not match the given input order. The authors of Harmony have developed an implementation of conflict detection as done in Aria, by localized modifications to PostgreSQL .[4] We use this scheme in our implementation.

## 9.5 Other Related Work

Ledger databases [51] provide verification and auditing support. LedgerDB [49] does not support SQL, whereas SQL Ledger [5] provides forward integrity for updates on an SQL database using a blockchain. However, these systems require an audit process [6] for detecting corruption, and meanwhile wrong answers may be returned. Further they do not provide any efficient means for recovery from corruption.

Protection of data in outsourced databases relies on only one copy of the database, with query results verified using integrity proofs or "verification objects" . All the existing results e.g. [11], [27], among many others, work only for very limited query types, and are also very expensive, making them impractical, unlike BFT replicated state machine approaches such as ours.

TAPIR [52] provides transactional guarantees on top of inconsistent replication. However, it does not deal with Byzantine failure.

## 9.6 Future work

In many applications, most of the transactions update a small amount of data. Hence, it is possible to scale the system further, by sharding the dataset so that reads and updates of most transactions can be satisfied at a single shard . Each shard may be hosted at a subset of cluster replicas instead of all. For transactions which will access or update multiple shards,

Partitioning the database enables to scale the system further. In Calvin [44] model, remote partition read stage is completed prior to executing transaction write stage. This model is applicable for our design as well.

Another possible model is where all reads are executed first and then all writes are committed. This limits the concurrency as the reads of next transaction in sequence depend on writes of the current transaction.

Raft and Kafka do impose overheads, and optimizing their performance to further improve throughput is an area of future work.

[4] Available at https://github.com/zllai/AriaBC.

# 10 Conclusion

We presented a design for protecting database data from corruption, based on the Byzantine-fault tolerant replicated state machine approach. As long as only a minority of replicas are compromised, our design supports recovery of the corrupt replicas while the system is fully operational, without any downtime. Transactions are executed strictly in their order as per input log, thereby not batching as in Aria [23] or Harmony [22] but we get some benefit with group commit in NuRaft [29] and Postgres. We show that the approach can be implemented quite efficiently, on top of PostgreSQL, with relatively small changes to PostgreSQL itself. The system provides throughput of over 11000 transactions per second, and recovery is shown to be very efficient.

We observe that signature verification, Merkle tree maintenance and support for determinism impose relatively small overheads on transaction processing throughput.

We believe that our approach will have significant practical impact, given the increased risks of attacks such as ransomware or advanced penetration attacks that compromise servers.